\documentclass[aps,prl,twocolumn,showpacs,superscriptaddress,groupedaddress]{revtex4}  

\usepackage{graphicx}  
\usepackage{dcolumn}   
\usepackage{bm}        
\usepackage{amssymb}   

\begin{document}



\author{Benjamin B Machta}
\email{bmachta@princeton.edu}
\affiliation{Lewis Sigler Institute, Princeton University, Princeton, NJ, 08544}
\affiliation{Department of Physics, Princeton University, Princeton, NJ, 08544}

\title{A dissipation bound for thermodynamic control}

\begin{abstract}
Biological and engineered systems operate by coupling function to the transfer of heat and/or particles down a thermal or chemical gradient. 
In idealized \textit{deterministically} driven systems, thermodynamic control can be exerted reversibly, with no entropy production, as long as the rate of the protocol is made slow compared to the equilibration time of the system.   Here we consider \textit{fully realizable, entropically driven} systems where the control parameters themselves obey rules that are reversible and that acquire directionality in time solely through dissipation.  We show that when such a system moves in a directed way through thermodynamic space, it must produce entropy that is on average larger than its generalized displacement as measured by the Fisher information metric.  This distance measure is sub-extensive but cannot be made small by slowing the rate of the protocol.

\end{abstract}
\maketitle


It is generally believed that the fundamental laws of physics preserve phase space volume and are almost the same in reverse~\footnote{To be fully symmetric under time reversal, the system must also undergo the application of operators CP, conjugation of charge and inversion of all spatial coordinates.  This detail is not thought to contribute meaningfully to the macroscopic breaking of time reversal under discussion here}.  However, the macroscopic world with which we interact is dominated by biological and engineered machines that are \textit{dissipative} and which strongly break time reversibility; although their microscopic components obey the phase space volume preserving laws of physics, they couple movement to the production of entropy, leading them to move more often than not in a particular direction.  An engine couples the movement of heat down a thermal gradient to mechanical motion, leading it to move forward more often than backwards.  The macroscopic engines ubiquitous in engineered systems are sufficiently large and dissipative as to rarely move in reverse.  However, the processive but stochastic molecular motors common in biology do indeed take frequent steps backwards~\cite{Mora09,Bormuth09,Kodera10,Uchihashi11,Jannasch13,Battle15}, reminding us of the entropic and necessarily stochastic underpinnings of their directed motion.



How much entropy must be produced to ensure that a thermodynamic system will, on average, move in a forward direction? Most previous analysis of this question has assumed that the system is driven through the action of some number of control parameters, $\lambda^\mu$, which couple to degrees of freedom, $\left\{x\right\}$, of a controlled system through a term in the Hamiltonian $-\lambda^\mu \Phi_\mu(\left\{x\right\})$, where $\Phi_\mu(\left\{x\right\})$ is the conjugate force to $\lambda^\mu$.  As control parameters are moved \textit{deterministically} through some predetermined protocol $\vec{\lambda}(t)$, they do a stochastic amount of work on the system $W=\int dt  \frac{d\lambda^\mu}{dt} \Phi_\mu(t)$ where stochasticity arises due to the probabilistic nature of $\Phi_\mu(t)$ which depend implicitly on the system's state, $\left\{x\right\}$.   For such deterministically driven systems, as the protocol is made infinitely slow, $\left<W\right>\rightarrow \Delta F$, and the protocol becomes dissipationless.  The aim of this paper is to present a new bound that applies to \textit{fully realizable, entropically driven} systems, where all components obey microscopic reversibility, acquiring directionality in time only through dissipation.  We show that controlling such a system entails a finite entropic cost that cannot be removed by lengthening the time of the protocol.

The last two decades have seen enormous progress on understanding deterministically driven systems, with surprising equalities applying to ensemble averages of entropy production~\cite{Jarzynski97} and statistical properties of microscopic trajectories~\cite{Crooks99}.  In addition, new bounds constrain the entropy production associated with finite time protocols. 
Using arguments from linear response, Sivak and Crooks~\cite{Sivak12} showed that any protocol starting at $\vec{\lambda}_0$ and ending at $\vec{\lambda}_f$, to be completed in time $t_{\text{max}}$ must dissipate at least 
\begin{equation}
\label{eq:Sivak12Bound}
\left< \Delta S_{\text{tot}} \right> \geq \frac{\tilde{\mathcal{L}}^2\left(\vec{\lambda}_0,\vec{\lambda}_f \right)}{t_{\text{max}}} + \mathcal{O}\left(\frac{1}{t_{\text{max}}^2}\right)
\end{equation}
 entropy (in units where Boltzmann's constant $k_B=1$).   Here $\tilde{\mathcal{L}} (\vec{\lambda}_0,\vec{\lambda}_f )$ is the geodesic distance between initial and final control parameters in the metric space where infinitesimal length $d\tilde{l}$ between $\lambda$ and $\lambda+d \lambda$ is given by $d\tilde{l}^2= \tilde{g}_{\mu\nu} \left(\lambda \right) d\lambda^\mu d\lambda^\nu$, where here and throughout repeated indeces are summed and $\tilde{g}$ is a `friction tensor' defined by
\begin{equation}
\label{eq:gtilde}
\tilde{g}_{\mu\nu} \left(\lambda \right)=\int\limits_0^\infty dt \left< \left(\Phi_\mu\left( 0 \right)-\left<\Phi_\mu\right>_\lambda\right) \left(\Phi_\nu\left( t \right)-\left<\Phi_\nu\right>_\lambda\right) \right>_\lambda.
\end{equation}
Crooks as well as Burbea and Rao~\cite{Crooks07,Burbea82} considered dissipation in related protocols in which thermodynamic parameters are moved in $N$ discrete steps, with the system coming fully to equilibrium between each step.  In this case the dissipation is bounded by 
\begin{equation}
\label{eq:Crooks07Bound}
\left< \Delta S_{\text{tot}} \right> \geq \frac{\mathcal{L}^2\left(\vec{\lambda}_0,\vec{\lambda}_f \right)}{n_{\text{steps}}} + \mathcal{O}\left(\frac{1}{n_{\text{steps}}^2}\right),
\end{equation}
where the tilde-free $\mathcal{L} (\vec{\lambda}_0,\vec{\lambda}_f )$ is the geodesic distance between initial and final control parameters in the metric space defined by the Fisher information matrix (FIM).   The FIM measures statistical distinguishability between nearby models~\cite{Ruppeiner95,Machta13} and in this parameterization takes the form
\begin{equation}
\label{eq:FIM}
g_{\mu\nu} \left(\lambda \right)= \left< \Phi_\mu\Phi_\nu \right>_\lambda- \left< \Phi_\mu\right>_\lambda \left< \Phi_\nu \right>_\lambda=\frac{\partial \left< \Phi_\mu\right>_\lambda}{\partial \lambda^\nu}.
\end{equation}

In each of these deterministically driven cases, the entropy production can be reduced to an arbitrarily small value by taking the limit where either $t_{\text{max}} \rightarrow \infty$ or $n_{\text{steps}} \rightarrow \infty$.  In this manuscript, we will consider fully realizable systems, where the control parameters themselves obey dynamics that are in accord with the principles of time reversal invariance, moving forward more often than backwards only due to the increased volume of phase space associated with movement in that direction.  In this \textit{entropically} driven limit, there is an intrinsic cost associated with thermodynamic control, which, \textit{even as the procedure is made arbitrarily slow and fine-grained} remains bounded by 
\begin{equation}
\label{eq:bound}
\left< \Delta S_{\text{tot}} \right> \geq 2 \mathcal{L}\left(\vec{\lambda}_0,\vec{\lambda}_f \right)
\end{equation}
where $\mathcal{L}$ is the geodesic distance in the same metric space of the FIM defined in Eq.~\ref{eq:FIM}.  Demonstrating this bound and discussing its implications are the subject of this paper.

\begin{figure}
	\includegraphics[width=\columnwidth]{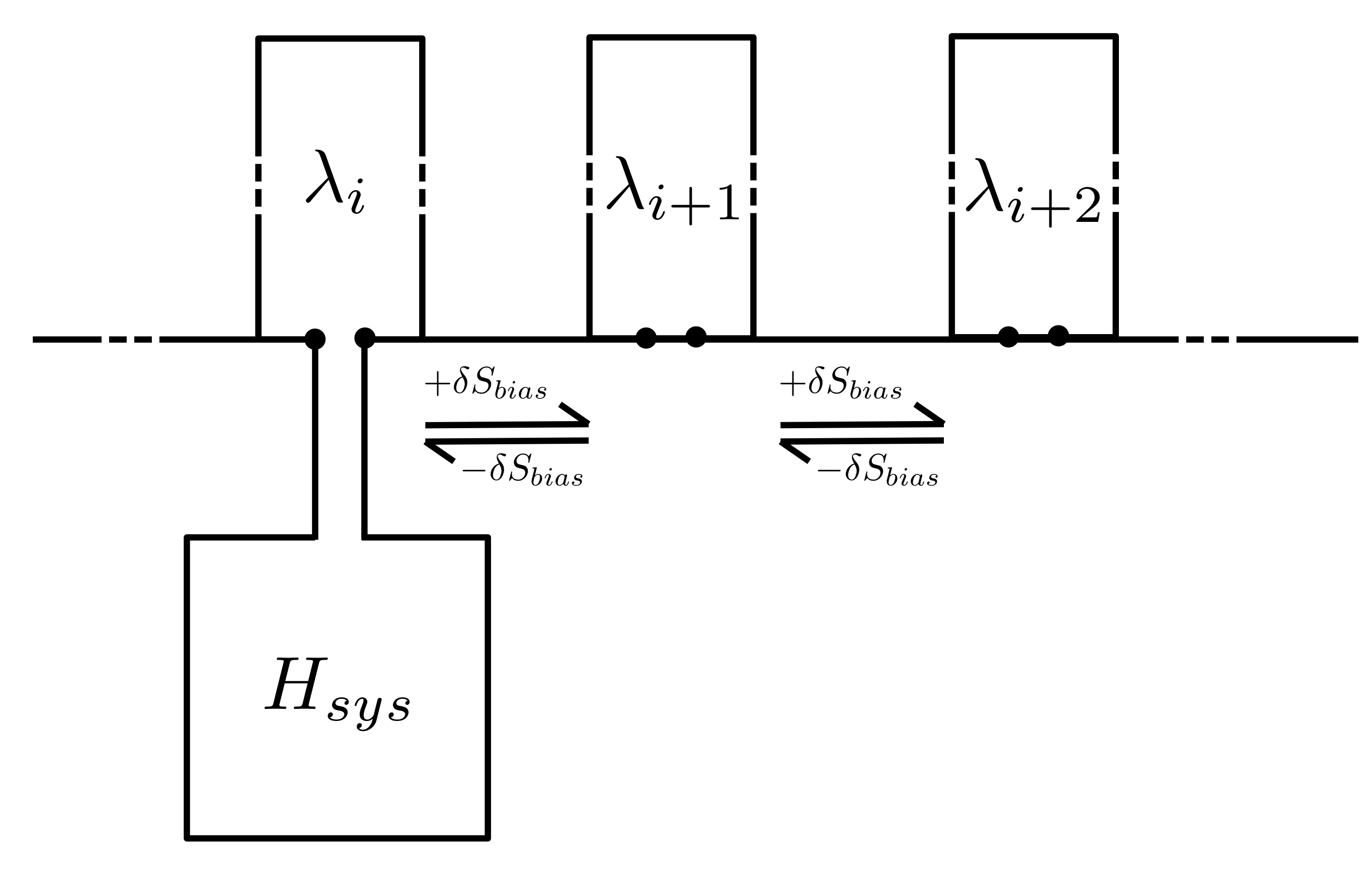}
	\caption{\label{fig:model} A fully realizable system consisting of a controlled system of interest with Hamiltonian $H_{\text{sys}}$, a series of particle reservoirs $i$ each with chemical potential $\lambda_i/\beta$, and an unspecified source of bias entropy $\delta S_{\text{bias}}$.  At a fast timescale the system exchanges particles with bath $i$, and at a much slower timescale it hops from bath $i$ to $i \pm 1$, producing $\pm \delta S_{\text{bias}}$ entropy.  When $\delta S_{\text{bias}}$ is small, the system tends to drift, hopping both forwards and backwards, during which it moves particles down their chemical gradient, dissipating additional entropy.  As argued in the text, average forward movement requires a minimum entropy production that cannot be removed either by making $d\lambda=\lambda_{i+1}-\lambda_{i}$ or $\delta S_{\text{bias}}$ small.  }
\end{figure}

Before looking at the more general case, consider a system with just a single control parameter, $\lambda$, the chemical potential of particles interacting through some Hamiltonian $H_{\text{sys}}(\left\{x\right\})$, where $\left\{x\right\}$ denotes the micro-state of the system (see Fig.~\ref{fig:model}). Particle number, $N(\left\{x\right\})$ is a function of $\left\{x\right\}$ and at time $t$ the system may exchange particles with particle reservoir $i(t)$ at chemical potential $\lambda_i/\beta$.  After reaching equilibrium with bath $i$, the probability of finding the controlled system in state $\left\{x \right\}$ is given by an appropriate Boltzmann distribution,
\begin{equation}
P_i\left(\left\{x\right\}\right) = \frac{\exp \left(-\beta H_{\text{sys}}(\left\{x\right\}) + \lambda_i N\left(\left\{x\right\}\right)\right)}{Z_i}.
\end{equation}
With these definitions, $\Omega=-\log(Z_i)/\beta $ is the grand potential, 
and the Helmholtz free energy 
is given by $F_{\text{sys},i} = \Omega+\lambda_i \left<N\right>_i /\beta$, where $\left< N \right>_i=\partial \log(Z_i) / \partial \lambda$.  The FIM of this system can be expressed as (see Eq.~\ref{eq:FIM})
$g=\left<N^2 \right>_i-\left<N\right>_i^2= \frac{\partial \left< N\right>_i }{ \partial \lambda}=-\frac{\partial^2 \Phi}{\partial \lambda^2}$
where the subscripts on $g$ are omitted as the system contains only a single parameter.  Changes in the Helmholtz free energy describe changes in the entropy of the system and its coupled heat bath. Additionally, in a process in which $\Delta N_i$ particles flow from the various particle reservoirs into the system, each particle reservoir's entropy changes by an amount 
$\Delta S_{i} =\lambda_i \Delta N_i$.

In addition to dynamics in which the system comes to equilibrium with bath $i$, our system may also disconnect from bath $i$ and reconnect to baths $i+1$ or $i-1$, that are held at $\lambda_{i\pm1} = \lambda_i \pm d\lambda$.  Let us first consider dynamics in which these steps happen in the forward and reverse directions with equal rates.  In this directionless steady state, the complete system consisting of baths and controlled system is \textit{not} in thermal equilibrium.  In fact, as it diffuses back and forth, the controlled system will tend to move particles down their concentration gradients.  Consider a sequence in which the system starts in equilibrium connected to reservoir $i$, hops to reservoir $i+1$, comes to equilibrium there, and then hops back to reservoir $i$ where it again comes to equilibrium.  At the end of this sequence, the entropy of the thermal bath and controlled system remain unchanged.  However, on average, particles have been transferred from bath $i+1$ to bath $i$;  during the forward hop the system carries, on average $\left< N \right>_{i}$ particles while in the backwards trajectory it carries $\left< N \right>_{i+1}$=$\left< N \right>_{i} +g d\lambda+\mathcal{O}(d\lambda)^2$ particles (see Eq.~\ref{eq:FIM}).  

In fact a hop in either direction produces, on average, total entropy $g \left(d\lambda\right)^2/2$.  Consider a sequence in which the systems hops from bath $i$ to bath $i \pm 1$, where it comes to equilibrium.  The Helmholtz free energy of the system changes by 
$\beta \Delta F_{\text{sys}}=\pm g \lambda_i d\lambda + (\frac{g}{2} +\frac{\lambda_i}{2} \frac{\partial^2 \left<N\right>_i}{\partial \lambda^2})\left(d\lambda\right)^2+\mathcal{O}(d\lambda)^3$.
The entropy of reservoir $i\pm 1$ also changes.  On average, $\left<N\right>_{i+1}-\left<N\right>_{i}$ particles flow out of it, changing its entropy by 
$\Delta S_{i \pm 1}=(\lambda_i \pm d\lambda_i )
(\pm g d\lambda+\frac{\lambda_i}{2} \frac{\partial^2 \left<N\right>_i}{\partial \lambda^2}(d\lambda)^2+\mathcal{O}(d\lambda)^3)$.
Thus the system+bath+reservoir entropy,
$
S_{\text{sys}+}= -\beta F+\sum_i S_{i}
$
increases, on average, by
\begin{equation}
\label{eq:sys+}
\left<\Delta S_{\text{sys}+}\right>_{i\rightarrow i \pm1}=\frac{g}{2}(d\lambda)^2 + \mathcal{O}(d\lambda)^3.
\end{equation}

So far this system has no well defined direction of operation, despite its dissipative nature. To bias towards movement in the forward direction, a step from one particle reservoir to the next may couple to the dissipation of an unspecified source of biasing entropy $\delta S_{\text{bias}}$, so that a step from $i \rightarrow i+1$ is more likely than one in the reverse direction, with the forward and reverse rates $r_{\pm}$, as well as the average number of forward and reverse hops, $n_\pm$ related by~\cite{Crooks99,Feng08}:
\begin{equation}
\label{eq:bias}
\left<n_-\right>=\left<n_+\right> \times \exp(-\delta S_{\text{bias}}). 
\end{equation}
Microscopically, $\delta S_{\text{bias}}$ could arise from many sources.  For example, forward movement might be accompanied by the transfer of a fixed amount of energy down a thermal gradient, or of particles down a chemical gradient.  It is important to note that the reverse step does indeed reduce the entropy of the world by $\delta S_{\text{bias}}$. 

Let us now consider the average dissipation associated with moving system parameters from $\lambda_0$ to $\lambda_f$, over a trajectory through which $g$ is constant so that $\mathcal{L} \left(\lambda_0,\lambda_f \right)=\sqrt{g}\left|\lambda_f-\lambda_0 \right|$.  The total average dissipation associated with this trajectory is given by the sum of bias related dissipations and dissipations associated with hops, $\Delta S_{\text{sys}+}$:
\begin{equation}
\label{eq:STot}
\left<\Delta S_{\text{tot}} \right>= \left<n_+-n_-\right> \delta S_{\text{bias}}+\left<n_+ + n_-\right> \left< \Delta S_{\text{sys}+}\right>
\end{equation}
In any trajectory which moves from $\lambda_0$ to $\lambda_f$,  $n_+-n_- =\left(\lambda_f - \lambda_0\right)/d\lambda$. Combining with Eqs.~\ref{eq:sys+}~and~\ref{eq:bias} the total average dissipation is given by
\begin{equation}
\left<\Delta S_{\text{tot}}\right> =\mathcal{L} \left(\lambda_0,\lambda_f \right) \left(\frac{\delta S_{\text{bias}}}{\sqrt{g (d\lambda)^2}} + \frac{\sqrt{g (d\lambda)^2}}{2 \tanh(\delta S_{\text{bias}}/2)} \right).
\end{equation}
An idealized system could have any stepsize $d\lambda$ and any $\delta S_{\text{bias}}$.  Optimizing over these to find the values that minimize the dissipation we find that
\begin{equation}
\begin{array}{rl}
g \left(d\lambda^{opt}\right)^2 & =  \left(\delta S_{\text{bias}}^{opt}\right)^2 \\ \\
\delta S^{opt}_{\text{bias}} & \rightarrow 0 \\ \\
\left<\Delta S^{opt}_{\text{tot}}\right> &=2\mathcal{L} \left(\lambda_0,\lambda_f \right),
\end{array}
\end{equation}
demonstrating Eq.~\ref{eq:bound} for this system. 

To achieve the bound, the system must be in the continuum limit where $d\lambda$ and $\delta S_{\text{bias}}$ are small.   Furthermore, if $\delta S_{\text{bias}}>\sqrt{g}d\lambda$ then entropy is produced primarily in breaking the time-reversal-invariance of the control parameters, though the work done on the system can be made arbitrarily close to the change in free energy.  On the other hand, if $\delta S_{\text{bias}}<\sqrt{g}d\lambda$ then most entropy is produced in futile cycles in which the system moves back and forth in $\lambda$ space.  Insight into the optimum parameters can be gained by considering the `proper' velocity and diffusion coefficient where proper distance is measured by the FIM (where for generality we have added back indeces on $g$) :
\begin{equation}
\label{eq:VandD}
\begin{array}{rl}
V &= \sqrt{g_{\mu\nu}\frac{\partial \left< \lambda^\mu \right>}{\partial t} \frac{\partial \left<\lambda^\nu \right>}{\partial t} }\\ \\
D &=\frac{1}{2}g_{\mu\nu} \frac{\partial \left[ \lambda^\mu \lambda^\nu \right]}{\partial t} 
\end{array}
\end{equation}
where square brackets denote a second cumulant.  It is important to note that these each have dimensions of inverse time since proper distance as measured by the FIM is dimensionless.  For this system, in the continuum limit, $V=\sqrt{g} r_+ d\lambda \delta S_{\text{bias}}$ while $D=g r_+ (d\lambda)^2$.  As such, the optimum occurs when $D=V$.


These results also hold for a generalized system with multiple control parameters $\lambda^\mu$, which could be baths of different particles, temperature, displacements or other quantities conjugate to a generalized force.  Taking a continuum trajectory, we assume that the controlled system will be moved through a one-dimensional track $\vec{\lambda}(\tau)$, with motion along the track consisting of (one dimensional) diffusion and drift, microscopically arising from dynamics analogous to those considered in the one dimensional example.  At a particular point $\vec{\lambda}(\tau)$, using the definitions provided in Eq.~\ref{eq:VandD}, 
$V=\frac{d \left<\tau\right>}{d t}\sqrt{g_{\mu\nu} \frac{\partial \lambda^\mu}{\partial \tau}\frac{\partial \lambda^\nu}{\partial \tau}}$ and 
$D=\frac{1}{2}\frac{d\left[ \tau^2 \right]}{d t} \left(g_{\mu\nu} \frac{\partial \lambda^\mu}{\partial \tau}\frac{\partial \lambda^\nu}{\partial \tau}\right)$.  Similar analysis shows that in this multi-parameter continuum system, entropy is produced at the following rate:
\begin{equation}
\begin{array}{rl}
\frac{d\left<S_{\text{sys}+}\right>}{dt} &= D \\ \\
\frac{d\left<S_{\text{bias}}\right>}{dt} &= V^2 / D \\ \\
\end{array}
\end{equation}
The average amount of time it takes to traverse a segment of length $d\tau$ is given by $\left<dt\right>=\sqrt{g_{\mu\nu}\frac{\partial \lambda^\mu}{\partial \tau}\frac{\partial \lambda^\nu}{\partial \tau}}d\tau/V$, so that the dissipation associated with movement along a segment of length $d\tau$ is given by $\left<dS_{\text{tot}}\right>= \left(D/V+V/D \right)\sqrt{g_{\mu\nu}\frac{\partial \lambda^\mu}{\partial \tau}\frac{\partial \lambda^\nu}{\partial \tau}}d\tau $.  This is minimized when $D^{opt}=V^{opt}$ so that the minimum average dissipation associated with a stochastic protocol that on average moves from $\vec{\lambda}_0$ to $\vec{\lambda}_f$ along path $\vec{\lambda}(\tau)$ is given by
\begin{equation}
\begin{array}{rl}
\left<\Delta S^{opt}_{\text{tot}}\right> &= 2 \int\limits_{\tau_0}^{\tau_f} \sqrt{g_{\mu\nu}\frac{\partial \lambda^\mu}{\partial \tau}\frac{\partial \lambda^\nu}{\partial \tau}}d\tau \\ \\
&\geq 2 \mathcal{L}\left(\vec{\lambda}_0,\vec{\lambda}_f \right). 
\end{array}
\end{equation}
Note that the first equality defines (twice) the path length.  Thus, the inequality below follows from the definition of $\mathcal{L}$ as the length of the minimum length (geodesic) path.  Equality is achieved only when $\vec{\lambda}(\tau)$ is the minimum length geodesic connecting $\vec{\lambda}_0$ with $\vec{\lambda}_f$.

It is interesting to compare the bound presented here with those found by Crooks and Sivak (Eqs.~\ref{eq:Sivak12Bound} and~\ref{eq:Crooks07Bound} and Refs. \cite{Crooks07,Sivak12}).  Each of these previous bounds roughly correspond to limits in which $\delta S_{\text{bias}}$ is infinite, leading to a deterministic protocol, but where the process leading to that bias is ignored as a source of dissipation.  Each of these previous bounds can be made small by lengthening the time or number of steps associated with a protocol.  By contrast, neither the time, nor the number of steps enter the bound presented here.  

 The bounds presented by Crooks and Sivak~\cite{Crooks07,Sivak12} also scale differently with system size.  Both the FIM defined in Eq.~\ref{eq:FIM} and the metric introduced in Ref.~\cite{Sivak12} and Eq.~\ref{eq:gtilde} are extensive, scaling linearly with system size in the thermodynamic limit.  However, geodesic lengths scale as the square root of the metric.  As the geodesic lengths are squared in Eqs.~\ref{eq:Sivak12Bound} and ~\ref{eq:Crooks07Bound}, these bounds are extensive in the precise sense that controlling two identical and uncorrelated systems in tandem requires twice the dissipation.  By contrast, the bound presented here scales as the square-root of the system size;  two identical systems limited by this bound could be controlled in tandem with only $\sqrt{2}$ as much dissipation.  Thus, this bound becomes less important in the thermodynamic limit - it is more likely to be relevant for molecular machines than for large engineered systems, where extensive though finite time bounds~\cite{Sivak12} are more likely to be relevant.



The bound presented here is not in contradiction with the impressive array of experiments verifying the Jarzynski equality and Crooks relations.  
In these experimental tests, a control parameter is typically manipulated in the nearly deterministic regime.  For example, a bead attached to DNA or RNA is manipulated according to a predetermined force protocol by a laser operating as an optical tweezer.  The work done by the bead on the nucleotide polymer has been shown to obey both the Jarzynski equality~\cite{Liphardt02} and Crooks relations~\cite{Collin05}.  It would seem that, as a corollary, these experiments verify the dissipationless adiabatic limit for quasi-deterministically driven systems.  However, in these experimental paradigms the relatively enormous dissipation associated with the laser itself is ignored.  In the framework of this paper, this limit corresponds to the case where $\delta S_{\text{bias}}$ is very large, but where its contribution is ignored. 

It remains to be seen whether this bound meaningfully constrains the operation of real biological systems.  
%
Biology often utilizes molecular scale motors for which sub-extensive contributions to energy expenditure could be relevant.  
In muscle fibers, contraction is initiated by the binding of calcium to myosin, which then hydrolyze ATP to perform mechanical work.  Calcium binding acts much as $\delta S_{bias}$ does here, and biology spends substantial energy maintaining a low intracellular calcium concentration in resting cells.  As would be required to approach this bound, many biological motors are able to work in reverse, synthesizing ATP when pulled too hard~\cite{Itoh04}.  However, muscles do not seem to take advantage of this property, hydrolyzing ATP during both extension and contraction under load, and operating far from the reversible regime~\cite{Ryschon97}.  Indeed, in many contexts other considerations like speed, reliability and the constraints of chemistry~\cite{Hopfield74} may matter more than energetic efficiency.   England recently used non-equilibrium arguments to estimate the energetic cost of cellular replication, arguing that the dissipation required to rapidly build stable molecules dominates a bacterium's energy budget~\cite{England13}. 
Intriguingly, biological systems do seem to seek out highly cooperative, nearly-critical states~\cite{Mora11}, in particular in their cell membrane~\cite{Veatch08}. These regions of thermodynamic space have, among other properties, a high negative Ricci curvature~\cite{Ruppeiner95}, implying `closeness' to a much larger proper volume of distinct membrane compositions and corresponding physical properties.  

Further work will shed light on the extent to which this, or analogous bounds, apply to a larger class of fully realizable systems that do not contain any unphysical, deterministic, elements.  
This bound would seem to apply to idealized logically reversible computing schemes~\cite{Bennet73} that have been used to argue against~\cite{Bennet82} an intrinsic energetic cost associated with the mechanical manipulations underlying computation~\cite{Landauer61}.
In addition this bound constrains the efficiency of finite size heat engines.  A Carnot cycle operating between temperatures $T_h$ and $T_c$ with pressures $P_2$ and $P_1$ bounding the high temperature portion would have to dissipate entropy greater than $\Delta S_{\text{tot}} \geq 4\sqrt{N}\log(P_2/P_1)+4\sqrt{15N}\log(T_h/T_c)$ in every cycle.  Note the $\sqrt{N}$ dependence of this dissipation-  in the adiabatic limit an idealized deterministic system would be dissipationless, producing no net entropy but transferring an extensive $Q_h=N T_h \log(P_2/P_1)$ heat out of a hot bath and performing $W=N (T_h-T_c) \log(P_2/P_1)$ work. 
For fully realizable systems, any directed change must be driven by a finite, though sub-extensive, production of entropy, even if change is made slowly.
\section{acknowledgements}

Thanks to Michael Abbott, William Bialek, Chase Broedersz, Srivatsan Chakram, Gavin Crooks, Jon Machta, Georg Rieckh, Pierre Ronceray, James Sethna, Joshua Shaevitz, Mark Transtrum and Ned Wingreen for many useful discussions and feedback.  This work was supported by a Lewis-Sigler Fellowship.

\end{document}